\documentclass[aps,floats,prd,twocolumn,nofootinbib,superscriptaddress,preprintnumbers]{revtex4-2}

\usepackage{amssymb}
\usepackage{amsmath}

\usepackage{graphicx}
\usepackage{graphics}
\usepackage[caption=false]{subfig}
\usepackage{dcolumn}
\usepackage{color}
\usepackage{fancyhdr} 
\usepackage{hyperref}

\usepackage[utf8]{inputenc}

\makeatletter
\renewcommand*{\fnum@figure}{{\normalfont\bfseries \figurename~\thefigure}}
\renewcommand*{\@caption@fignum@sep}{\textbf{ : }}
\makeatother

\makeatletter
\renewcommand*{\fnum@table}{{\normalfont\bfseries \tablename~\thetable}}
\makeatother

\def\VEV#1{\left\langle #1 \right\rangle}

    \newcommand{\be}{\begin{equation}}
  \newcommand{\ee}{\end{equation}}
    \newcommand{\ba}{\begin{align}}
  \newcommand{\ea}{\end{align}}

\begin{document}

\author{Ranjan Laha
} 
\affiliation{Theoretical Physics Department, CERN, 1211 Geneva, Switzerland}
\author{Julian B.~Mu\~noz
} 
\affiliation{Department of Physics, Harvard University, Cambridge, MA 02138, USA}
\author{Tracy R.~Slatyer
} 
\affiliation{Center for Theoretical Physics, Massachusetts Institute of Technology, Cambridge, MA 02139, USA}

\title{INTEGRAL constraints on primordial black holes and particle dark matter}

\date{\today}

\begin{abstract}
The International Gamma-Ray Astrophysics Laboratory (INTEGRAL) satellite has yielded unprecedented measurements of the soft gamma-ray spectrum of our Galaxy.
Here we use those measurements to set constraints on dark matter (DM) that decays or annihilates into photons with energies $E\approx 0.02-2$ MeV.
First, we revisit the constraints on particle DM that decays or annihilates to photon pairs.
In particular, for decaying DM, we find that previous limits were overstated by roughly an order of magnitude.
Our new, conservative analysis finds that the DM lifetime must satisfy $\tau\gtrsim 5\times 10^{26}\,{\rm s}\times (m_{\chi}/\rm MeV)^{-1}$ for  DM masses $m_{\chi}=0.054-3.6$ MeV.
For MeV-scale DM that annihilates into photons INTEGRAL sets the strongest constraints to date.
Second, we target ultralight primordial black holes (PBHs) through their Hawking radiation.
This makes them appear as decaying DM with a photon spectrum peaking at $E\approx 5.77/(8\pi G M_{\rm PBH})$, for a PBH of mass $M_{\rm PBH}$.
We use the INTEGRAL data to demonstrate that, at 95\% C.L., PBHs with masses less than $1.2\times 10^{17}$ g cannot comprise all of the DM, setting the tightest bound to date on ultralight PBHs.
\end{abstract}

\maketitle
\preprint{CERN-TH-2020-056, MIT-CTP/5193}

\section{Introduction}

Dark matter (DM) is omnipresent in the universe, from sub-Galactic scales to galaxy clusters.
Despite its abundance, the nature of DM remains mysterious, as it has evaded all nongravitational direct and indirect probes thus far~\cite{Bertone:2004pz,Slatyer:2017sev,Lin:2019uvt}.  Given the enormous range in masses and interaction strengths of possible DM candidates, it is imperative to find new ways to probe different parts of their parameter space.

A powerful window into the nature of particle DM comes from its possible decay or annihilation into Standard Model (SM) particles.
In particular, gamma rays produced in DM interactions can be detected in our telescopes.
Here we will use the data from the  spectrometer (SPI) on board the International Gamma-ray Astrophysics Laboratory (INTEGRAL) satellite, covering the $E\sim 0.02-2$ MeV energy band.
By requiring that the DM emission does not exceed the Galactic diffuse flux measured by INTEGRAL~\cite{Bouchet:2008rp,Bouchet:2011fn}, we will set conservative bounds on various DM interactions, including revisiting some the constraints from Ref.~\cite{Essig:2013goa}.
In that work, limits were obtained using the Galactic-center gamma-ray spectrum from INTEGRAL~\cite{Bouchet:2008rp}.
That spectrum, however, only corresponds to the emission correlated with a fiducial template -- a combination of dust and CO maps -- and thus may not include photons from DM interactions, as the latter have a different spatial distribution.
Here, instead, we use the full emission profile from Ref.~\cite{Bouchet:2011fn}, which does not assume any spatial morphology.
By taking conservative assumptions, we find limits on the DM lifetime that are weaker by a factor of $\sim$ $3-10$ than found in Ref.~\cite{Essig:2013goa}.
For DM annihilating to electron-positron pairs our limits are broadly similar to Ref.~\cite{Essig:2013goa},
and we derive new limits on DM annihilating to photons.

We also search for gamma-ray emission from ultralight primordial black holes (PBHs).
PBHs, formed from SM plasma that collapsed due to its own gravity in the very early universe, are a possible solution to the DM puzzle~\cite{1966AZh....43..758Z,Hawking:1971ei,Carr:1974nx,Chapline:1975ojl,Meszaros:1975ef,Carr:1975qj,Sasaki:2018dmp}.
The fraction $f_{\rm PBH}$ of DM in the form of PBHs is constrained to be below unity for PBH masses $M_{\rm PBH}\gtrsim 10^{23}$ g ($\approx 5 \times 10^{-11} M_\odot$) via various observations, such as gravitational lensing~\cite{Tisserand:2006zx,Allsman:2000kg,Mediavilla:2009um,Calchi_Novati_2013,Griest:2013esa,Niikura:2017zjd,Carr:2017jsz,Zumalacarregui:2017qqd,Niikura:2019kqi,Sugiyama:2019dgt,Smyth:2019whb}, stellar dynamics~\cite{Yoo:2003fr,Quinn:2009zg,Monroy-Rodriguez:2014ula,Brandt:2016aco,Koushiappas:2017chw,Zhu:2017plg,Stegmann:2019wyz}, gravitational waves~\cite{Bird:2016dcv,Nakamura:1997sm,Kovetz:2017rvv,Ali-Haimoud:2017rtz,Clesse:2016vqa,Sasaki:2016jop,Authors:2019qbw,Raidal:2018bbj,Kavanagh:2018ggo,Vaskonen:2019jpv}, and the cosmic microwave background (CMB)~\cite{Ali-Haimoud:2016mbv,Blum:2016cjs,Poulin:2017bwe,Nakama:2017xvq}.
The situation is different for lower-mass PBHs, as their gravitational signatures are not strong enough to cause a measurable effect in existing data.
Instead, a promising avenue consists of searching for the radiation from ultralight PBHs as they Hawking evaporate~\cite{Hawking:1974sw,Hawking:1974rv,Page:1976df,Page:1976ki,Page:1977um}.
Previous works have searched for Hawking-evaporating PBHs through electron-positron pairs~\cite{Boudaud:2018hqb,1980AA....81..263O, okeke1980primary,1991ApJ...371..447M,Bambi:2008kx,DeRocco:2019fjq,Laha:2019ssq,Dasgupta:2019cae}, extra-Galactic gamma rays~\cite{Arbey:2019vqx,Ballesteros:2019exr,Carr:2020gox}, and the CMB~\cite{Poulter:2019ooo,Stocker:2018avm,Acharya:2019xla,Acharya:2020jbv}.
These studies have ruled out PBHs as the entirety of the DM for $M_{\rm PBH} \lesssim 10^{17}$ g.
Here we search for gamma rays from PBHs in the DM halo around the Milky Way (MW), which would appear as a modified blackbody spectrum in the INTEGRAL data.
We are able to rule out PBHs as the sole component of DM for $M_{\rm PBH} \lesssim 1.2 \times 10^{17}$ g, at 95\% C.L., setting the strongest bound to date on the mass of ultralight PBH DM.

This paper is structured as follows.
We begin in Sec.~\ref{sec:emission} by reviewing the gamma-ray emission from different DM models. 
We introduce the INTEGRAL data in Sec.~\ref{sec:data}, and use it in Sec.~\ref{sec:results} to find our constraints.
We conclude in Sec.~\ref{sec:conclusions}.

\section{Emission from Different DM Models}
\label{sec:emission}

Assuming a DM candidate of mass $M_{\rm DM}$, the differential gamma-ray flux produced by its decay or annihilation (denoted by $\alpha=1$ and 2, respectively) is
\be
\dfrac{d\Phi}{dE} = \dfrac{1}{2^{\alpha-1}} \dfrac{r_\odot}{4\pi} \left(\dfrac{\rho_\odot}{M_{\rm DM}} \right) \dfrac{J_{D/A}}{\Delta \Omega} \dfrac{dN}{dE dt}\,,
\ee
where $r_\odot=8.3$ kpc and $\rho_\odot=0.01\,M_\odot\,\rm pc^{-3}$ ($=0.4$ GeV cm$^{-3}$) are the Galacto-centric distance of the Sun and the local DM density, respectively~\cite{Abuter:2020dou,Buch:2018qdr,Schutz:2017tfp}.
If a different DM density is assumed, our constraints can be simply rescaled accordingly.
The differential flux formula can be neatly divided into a component that depends on the spatial distribution of DM, the $J_{D/A}$ factor ($D$ for decay and $A$ for annihilation), and one that depends on the photon spectrum per decay/annihilation multiplied by the rate at the Solar radius, $dN/(dE \, dt)$.
The former will not vary between models, so let us begin by describing it.

We will assume that the DM in the Milky-Way (MW) halo follows a Navarro-Frenk-White (NFW) profile~\cite{Navarro:1996gj}, as both MeV-scale particle DM and ultralight PBHs behave as cold DM (see Appendix~\ref{sec:appendix} for results assuming other profiles).  
We take a scale radius $r_h=17$ kpc for the MW halo~\cite{McMillan:2011wd}.
The differential $J_{D/A}$ factor, integrating along the line of sight, is
\be
\dfrac{dJ_{D/A}}{d\Omega} (\hat n) = \int \dfrac{ds}{r_\odot} \left(\dfrac{\rho(s, \hat n)}{\rho_\odot} \right)^\alpha\,,
\label{eq:dJ}
\ee
where $\rho$ is the DM density in the MW halo.
The details of the DM profile are not critical for evaporating PBHs or decaying DM, whereas they have a bigger impact for annihilating DM, although our results can be rescaled to other DM profiles.

The $J_{D/A}$ factor is calculated by integrating Eq.~\eqref{eq:dJ} over the patch of the sky considered, with an angular extension $\Delta \Omega$.
The observed flux over an energy band spanning the range $(E_1,E_2)$ is
\be
\Phi = \int_{E_1}^{E_2} \dfrac{d\Phi}{dE} dE\,,
\ee
with units of $\rm cm^{-2} s^{-1} sr^{-1}$.
This is the quantity we will compare with INTEGRAL observations of our Galaxy.

\begin{figure}[hbtp!]
	\centering
	\includegraphics[width=0.48\textwidth]{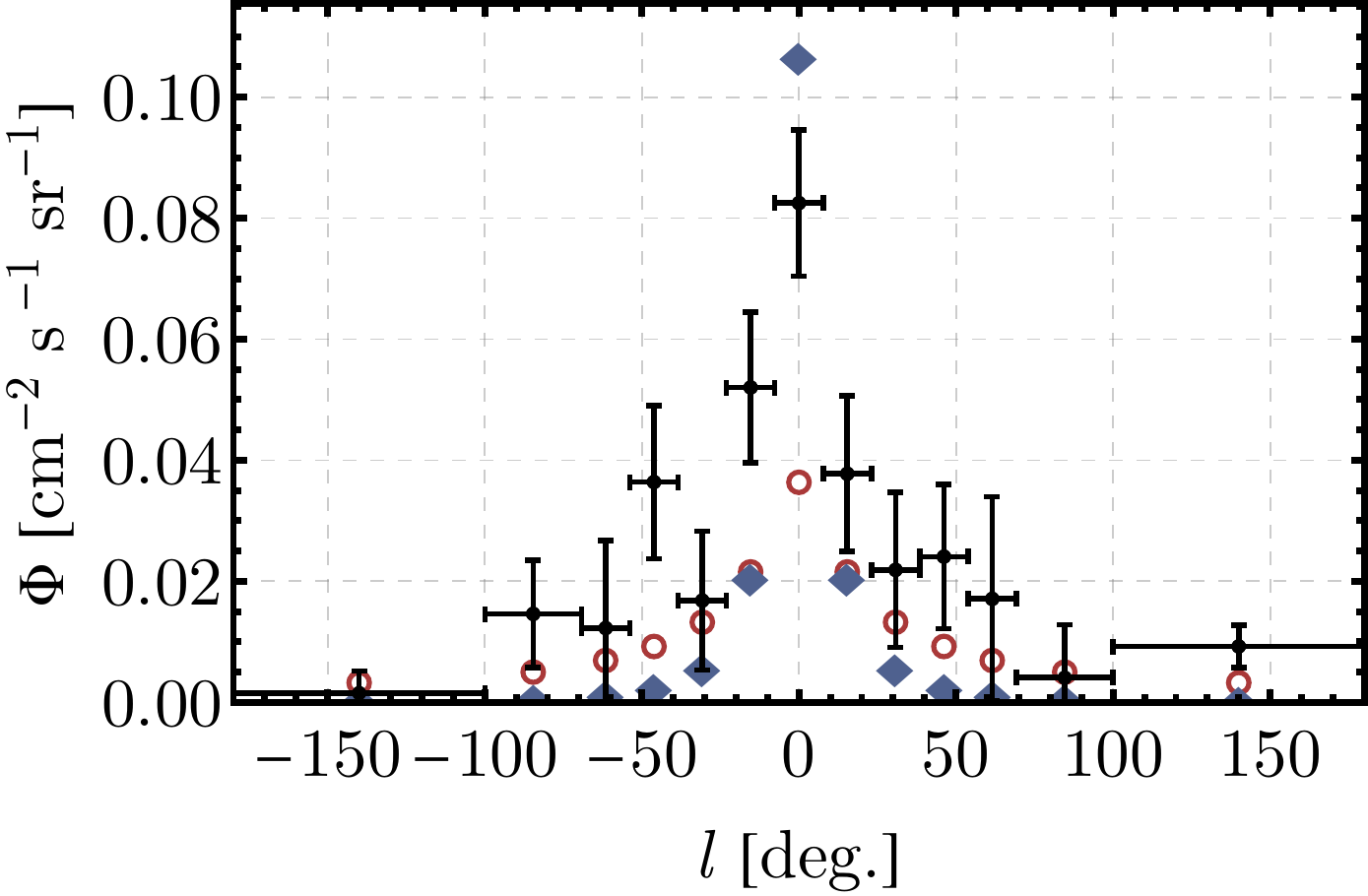}
	\includegraphics[width=0.48\textwidth]{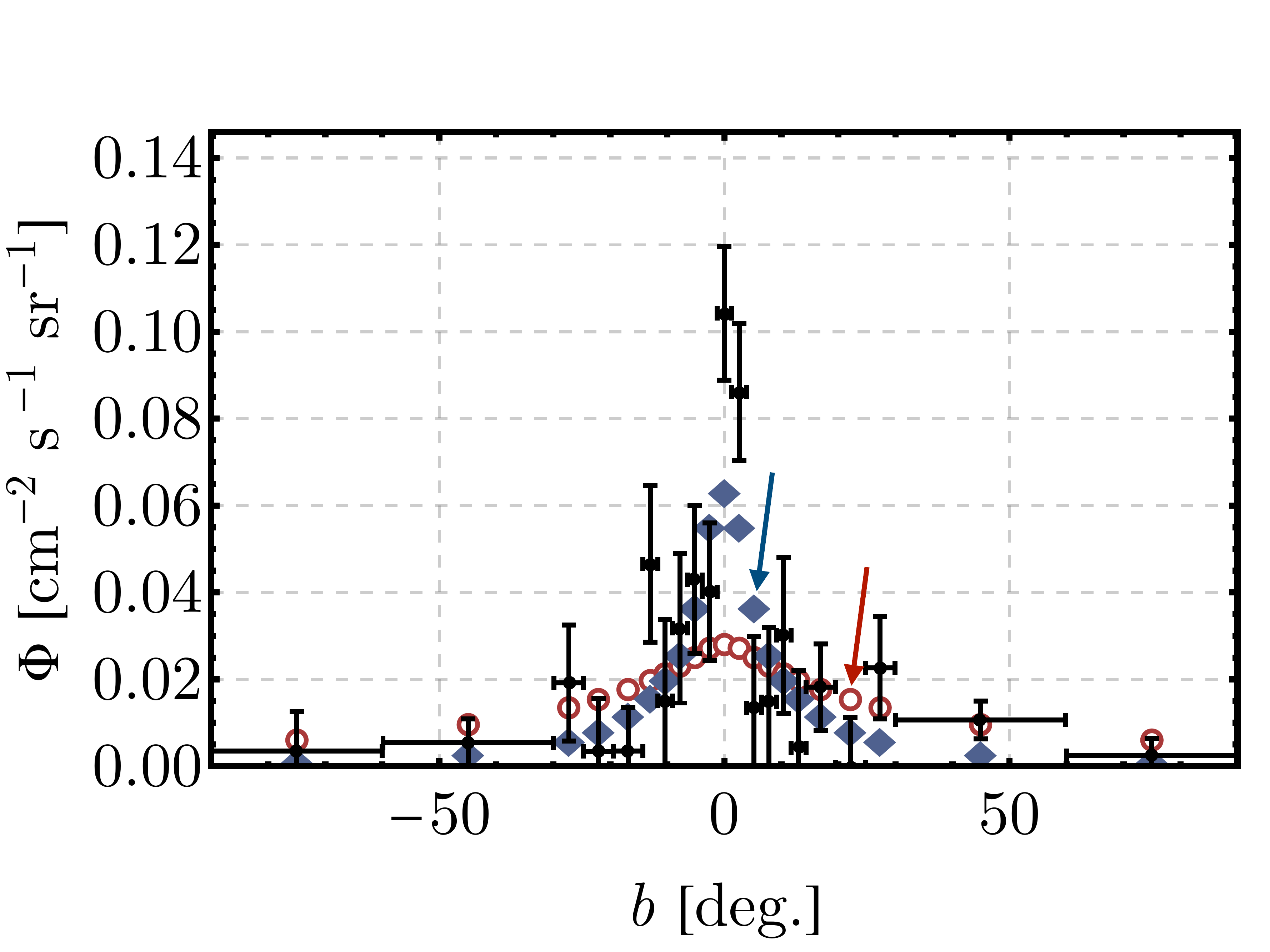}
	\caption{Galactic gamma-ray flux in the $0.2-0.6$ MeV energy band.
	The black crosses show the INTEGRAL measurements as a function of Galactic longitude $l$ ({\bf top}, integrated over $|b|<6.5$ deg.) and latitude $b$ ({\bf bottom}, integrated over $|l|<23.1$ deg.)~\cite{Bouchet:2011fn}.
	The red circles show the predicted emission from Hawking-evaporating PBHs with $M_{\rm PBH}=1.5\times 10^{17}$ g, if those composed the entirety of the DM in the MW,
	whereas the blue diamonds correspond to DM of mass $m_\chi = 0.5$ MeV annihilating to photon pairs with $\VEV{\sigma v}_{\gamma \gamma}=6 \times 10^{-31}$cm$^3$ s$^{-1}$.
	The two arrows in the bottom plot indicate the latitude bins where
    the DM emission exceeds the 1-$\sigma$ measured flux, for each model.
	}	
	\label{fig:fluxprofiles}
\end{figure}

\subsection{Particle Dark Matter}

We now turn to describe the emission spectrum in each of our models, beginning with the simpler cases of decaying and annihilating particle DM with mass $m_\chi$.

We first study DM decaying into two photons of energy $E$, where the emission spectrum from a DM particle with lifetime $\tau$ is
\be
\dfrac{dN}{dt dE} = \dfrac{2}{\tau} \,\delta_D\left(E-\dfrac{m_\chi}{2}\right)\,.
\ee
Another possible channel would consist of DM decaying to electron-positron pairs plus final-state radiation (FSR).
In this case, however, the CMB can set stronger constraints than Galactic observations~\cite{Slatyer:2016qyl}, so we do not consider it here.

For annihilating DM, we will focus on the case of 
DM annihilating to two photons, for which
\be
\dfrac{dN}{dt dE} = 2 \VEV{\sigma v}_{\gamma \gamma} \left(\dfrac{\rho_\odot}{m_\chi} \right) \,\delta_D\left(E-m_\chi\right)\,,
\ee
where $\VEV{\sigma v}$ denotes the thermally averaged annihilation cross sections.
We show, in Fig.~\ref{fig:fluxprofiles}, the Galactic flux profile in the $0.2-0.6$ MeV band, for DM annihilating to two photons, with $m_\chi = 0.5$ MeV and $\VEV{\sigma v}_{\gamma\gamma} = 6\times 10^{-31}$ cm$^3$ s$^{-1}$.
Halo substructure can cause an enhancement in the annihilating signal\,\cite{Ando:2019xlm}, which we conservatively ignore in our analysis.

\subsection{Ultralight PBHs}

Black holes (BHs), with mass $M_{\rm BH}$, evaporate over time, emitting particles roughly as a blackbody with temperature~\cite{Hawking:1982dh}
\be
T_{\rm BH} = \dfrac{1}{8\pi G M_{\rm BH}}\,.
\ee
For reference, $M_{\rm BH} = 10^{17}$ g corresponds to a BH temperature $T_{\rm BH} \approx 0.1$ MeV,
and thus the Hawking emission from these BHs will predominantly consist of neutrinos and photons.
We will focus on the latter, given the well-measured diffuse gamma-ray emission from the MW by INTEGRAL.

The spectrum of particles emitted from an evaporating BH does not exactly follow a blackbody distribution~\cite{Page:1976df,Page:1976ki,Page:1977um}.
In particular, the photon spectrum is given by~\cite{MacGibbon:1990zk}
\be
\dfrac{dN}{dt dE} = \dfrac{1}{2\pi} \dfrac{\Gamma(E,M_{\rm BH})}{e^{E/T_{\rm BH}}-1}\,,
\ee
where $\Gamma(E,M_{\rm BH}) = E^2 \sigma(E,M_{\rm BH})/\pi$ is the graybody factor, which accounts for the departure from pure blackbody emission, and $\sigma(E,M_{\rm BH})$ is the absorption cross section for spin-1 particles (such as photons).
At high energies, this cross section approaches the geometric limit, $\sigma(E \gg T_{\rm BH},M_{\rm BH}) \approx 27 \pi G^2 M_{\rm BH}^2$. 
Nonetheless, for $E\lesssim T_{\rm BH}$ this cross section is significantly lower, reducing the overall amount of BH emission.
For spin-1 particles, this increases the energy peak to $E\approx 5.77\,T_{\rm BH}$\,\cite{MacGibbon:2007yq}.
In this work we will focus on Schwarzschild BHs (as the emission is slightly different for spinning or charged BHs~\cite{Kerner:2008qv,Jiang:2005ba}), and we use the public code {\tt BlackHawk}~\cite{Arbey:2019mbc}\footnote{\url{https://blackhawk.hepforge.org/}} to compute the Hawking emission.

We show the flux for PBH DM with $M_{\rm PBH}=1.5\times 10^{17}$ g in the $0.2-0.6$ MeV band---where it peaks---in Fig.~\ref{fig:fluxprofiles}.
The Galactic emission from annihilating DM is more concentrated towards the Galactic Center, as opposed to that from PBHs (or generic decaying DM).

\section{INTEGRAL data}
\label{sec:data}

In order to constrain the emission from different DM models, we will use data from the SPI instrument on the INTEGRAL satellite, which roughly covers the $0.02-8$ MeV energy band\,\cite{2003A&A...411L..63V}.
In particular, we will employ the measurements of diffuse Galactic emission from Ref.~\cite{Bouchet:2011fn}, where point sources are simultaneously subtracted.
The coded-mask system of SPI is, in principle, only sensitive to differences in flux, and thus cannot observe the isotropic extra-Galactic background~\cite{Bouchet:2011fn}.
Nonetheless, some unknown amount of background radiation can appear as part of the INTEGRAL measurements, and this uncertainty dominates the error budget (and it is expected to be behind the $\sim 30\%$ increase in flux with respect to earlier INTEGRAL data~\cite{Bouchet:2008rp,Bouchet:PrivComm}).

In Refs.~\cite{Bouchet:2008rp,Bouchet:2011fn} the INTEGRAL data is reduced in two different ways.  The first way finds the spectrum of the Galactic inner radian, with a relatively fine energy resolution.
While this dataset might be optimal for constraining DM, especially for decays or annihilations to a photon line, the Galactic emission in this analysis is assumed to follow the morphology of a predetermined map~\cite{Bouchet:2008rp,Bouchet:2011fn}, given by a combination of dust, CO, and inverse-Compton emission, and not that due to DM.
Therefore, a DM signal may be hidden behind the projection onto these maps, as they are not guaranteed to account for every photon.
This was the dataset used to obtain constraints in Ref.~\cite{Essig:2013goa}.

The second is the profile of diffuse emission as a function of Galactic angle.
This is the dataset that we will use, as it does not assume any emission morphology, and thus  can be used to set conservative bounds.
On the one hand, the energy bands in this analysis are wider, which makes it harder to detect lines originating from DM.
On the other hand, the data at angles beyond the Galactic inner radian provides more constraining power, as astrophysical backgrounds can be more concentrated than the signal for decaying DM.

The INTEGRAL measurements are divided into five energy bands, with cuts at $E=0.027, 0.049, 0.1, 0.2, 0.6$ and 1.8 MeV.
We show the data in the fourth band ($E=0.2-0.6$ MeV) in Fig.~\ref{fig:fluxprofiles}, corresponding to the peak of emission for PBHs with $M_{\rm PBH} = 1.5\times 10^{17}$ g, and for DM annihilating to photons with $m_\chi=0.5$ MeV.
The latitude profiles are integrated over longitudes $|l|<23.1$ deg., whereas the longitude profiles are integrated over $|b|<6.5$ deg. (except the highest-energy bin, which has $|l|<60$ deg. and $|b|<8.2$ deg.).
By comparing with the emission from PBH DM, we see in Fig.~\ref{fig:fluxprofiles} that the constraints will be driven by intermediate latitudes, as opposed to the case of annihilating DM, which peaks closer to the Galactic Center.
Additionally, note that the data point at $b = 20-30$ deg. is below the expected emission from PBHs, given our chosen mass.
Likewise, the emission from annihilating DM with $\VEV{\sigma v}_{\gamma \gamma}=6 \times 10^{-31}$ cm$^3$ s$^{-1}$ is above the INTEGRAL data point at $b = 6-9$ deg.
Therefore these two models will be excluded at 68\% C.L.

\section{Results}
\label{sec:results}

We will now present conservative constraints on the DM models by requiring that their emission is below the maximum allowed by the INTEGRAL data.
The INTEGRAL error budget is not Poissonian, but is dominated by the fitting procedure, which simultaneously removes point sources and an isotropic extra-Galactic background. Likewise, errors between different energy bins are likely correlated.
Nevertheless, we will multiply by two the error bars reported in Ref.~\cite{Bouchet:2011fn} to obtain the fluxes at 95\% C.L.~at each energy bin.
Then, we will obtain our limits by requiring that the emission from each DM candidate is smaller than this 95\% C.L. flux.
We note that older observational data from the INTEGRAL satellite were used to constrain particle DM in Refs.\,\cite{Yuksel:2007xh, Mack:2008wu, Boyarsky:2007ge}.

\begin{figure}[hbtp!]
	\centering
	\includegraphics[width=0.48\textwidth]{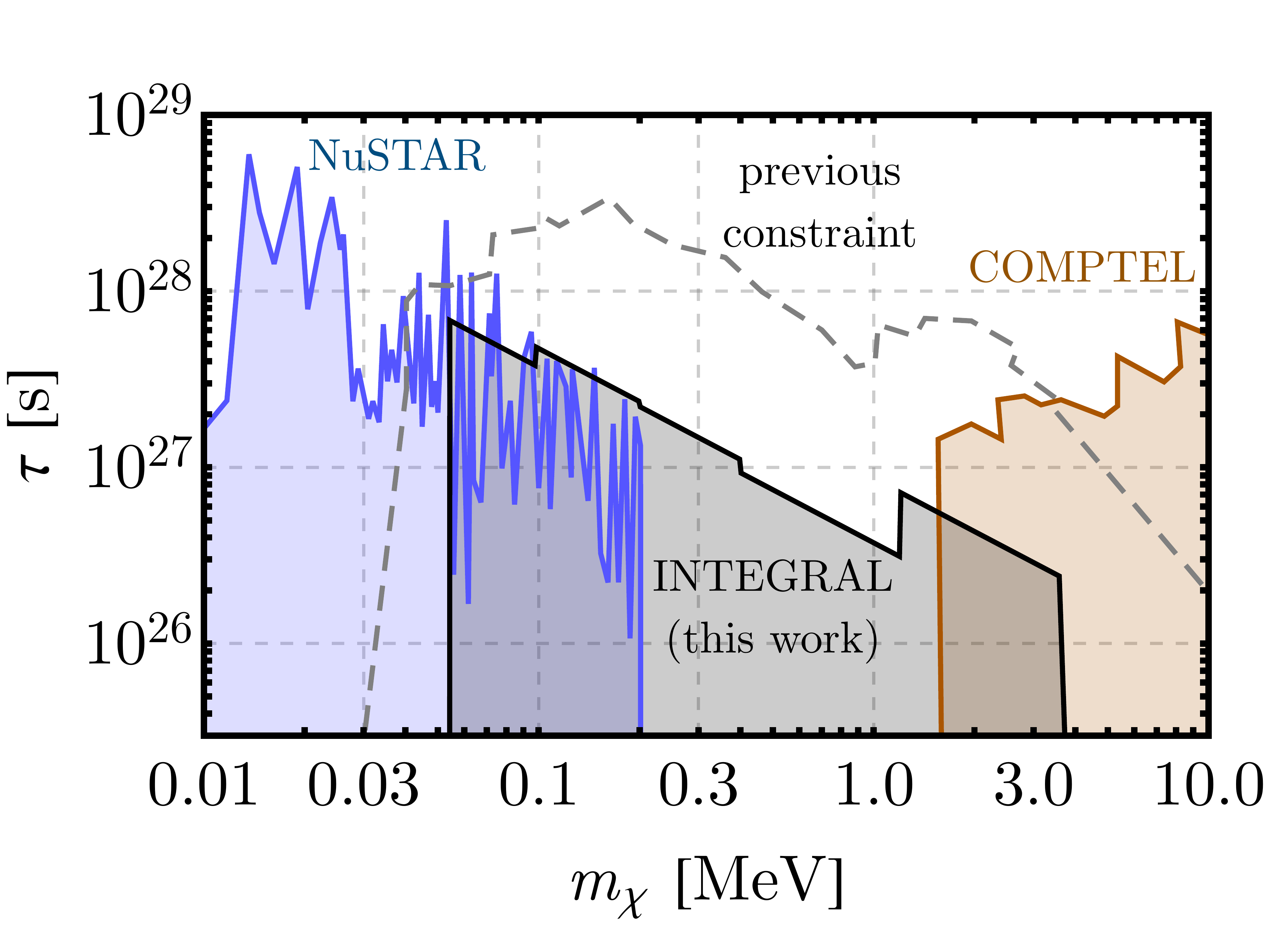}
	\caption{Constraints on the lifetime of decaying DM, $\tau$, assuming decay to two photons, as a function of its mass $m_\chi$.
	The orange and blue shaded regions are constrained by X-ray and gamma-ray data from COMPTEL~\cite{Essig:2013goa}, and NuSTAR~\cite{Ng:2019gch}.
	Our conservative re-analysis of INTEGRAL data yields the 95\% C.L.~constraints shaded in black, to be compared with the previous result from Ref.~\cite{Essig:2013goa} as
	 the dashed gray line.
	 The kinks in our limit (as well as those presented in Fig.\,\ref{fig:annDM_phot}) reflect the energy binning in the INTEGRAL data.
	}	
	\label{fig:tauconst}
\end{figure}

\subsection{Particle Dark Matter}

We begin by revisiting the INTEGRAL constraints on decaying particle DM from Ref.~\cite{Essig:2013goa}.

We show our constraint for decaying DM on Fig.~\ref{fig:tauconst}, along with the previous result of Ref.~\cite{Essig:2013goa}.
Our robust analysis weakens the INTEGRAL constraints by nearly an order of magnitude.
That is partially because not every photon is included in the data used in Ref.~\cite{Essig:2013goa}, as well as due to the loss of energy resolution (which would significantly help in this case).
Moreover, we do not include extra-Galactic photons from decaying DM, as those are not accounted for in the INTEGRAL/SPI data set that we use, which narrows the mass range that can be constrained.
We can probe DM masses $m_\chi \in [0.054-3.6]$ MeV, over which our constraint, in Fig.~\ref{fig:tauconst}, can be approximated by $\tau\gtrsim 5\times 10^{26}\,{\rm s} \times (m_{\chi}/\rm MeV)^{-1}$. 
Even when accounting for the weakening of the INTEGRAL limits, these are still 3 orders of magnitude stronger than those obtained from the CMB~\cite{Slatyer:2016qyl}.

\begin{figure}[hbtp!]
	\centering
	\includegraphics[width=0.48\textwidth]{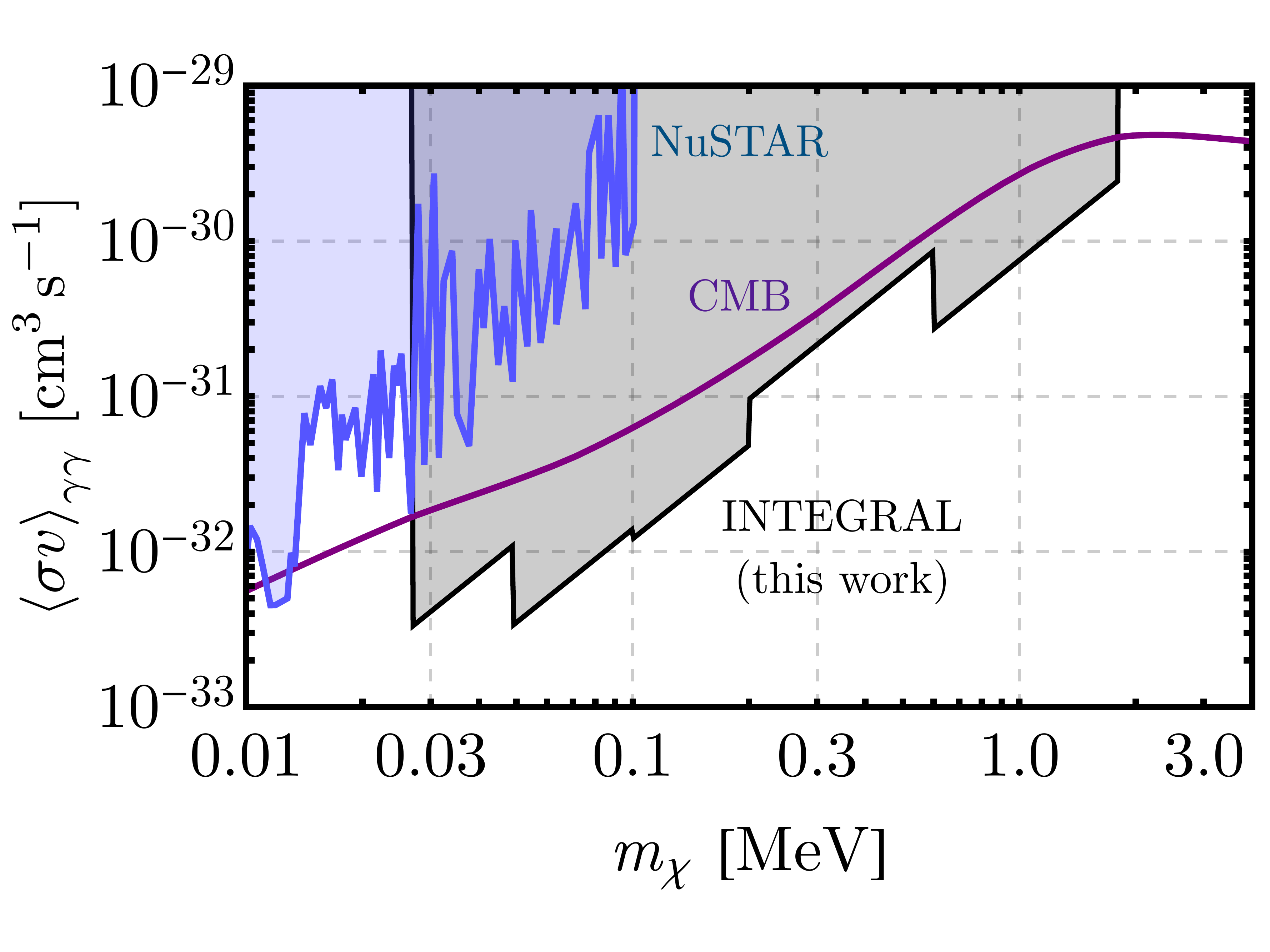}
	\caption{Constraints on the thermally averaged cross section of DM annihilating to two photons, $\langle \sigma v \rangle_{\gamma \gamma}$, as a function of its mass, $m_\chi$, for our INTEGRAL reanalysis (95\% C.L., in black),
	compared to the CMB $s$-wave limits~\cite{Slatyer:2015jla} (in purple), as well as the current best limits from NuSTAR~\cite{Ng:2019gch} (in blue).
	}
	\label{fig:annDM_phot}
\end{figure}

We now study the case of annihilating dark matter.
We show the limits for DM annihilating to two photons in Fig.~\ref{fig:annDM_phot}.
In this case the INTEGRAL data provides stronger constraints than the CMB~\cite{Slatyer:2016qyl} and NuSTAR observations of M31~\cite{Ng:2019gch}. 
While the NuSTAR constraints could be potentially extended to higher masses using Galactic-center data~\cite{Perez:2016tcq}, they will be less constraining than our results, so we do not show them here. 
All these constraints are significantly smaller than the thermal-relic cross section, and thus do not allow for a thermal relic within this mass range annihilating exclusively to two photons.

Additionally, DM with masses above the electron mass can annihilate to electron-positron pairs plus FSR.
We note, in passing, that for this case we find good agreement with the constraints of Ref.~\cite{Essig:2013goa}, as the FSR spectrum is fairly broad, so the loss of energy resolution in our analysis does not change the results significantly.

\subsection{Ultralight PBHs}

We follow the same approach for the PBHs, except now we will phrase our constraints in terms of the maximum fraction $f_{\rm PBH}$ of the DM that is allowed to be comprised of PBHs, assuming a monochromatic mass function, and an NFW distribution for PBHs, even for low values of $f_{\rm PBH}$.

\begin{figure}[btp!]
	\centering
	\includegraphics[width=0.48\textwidth]{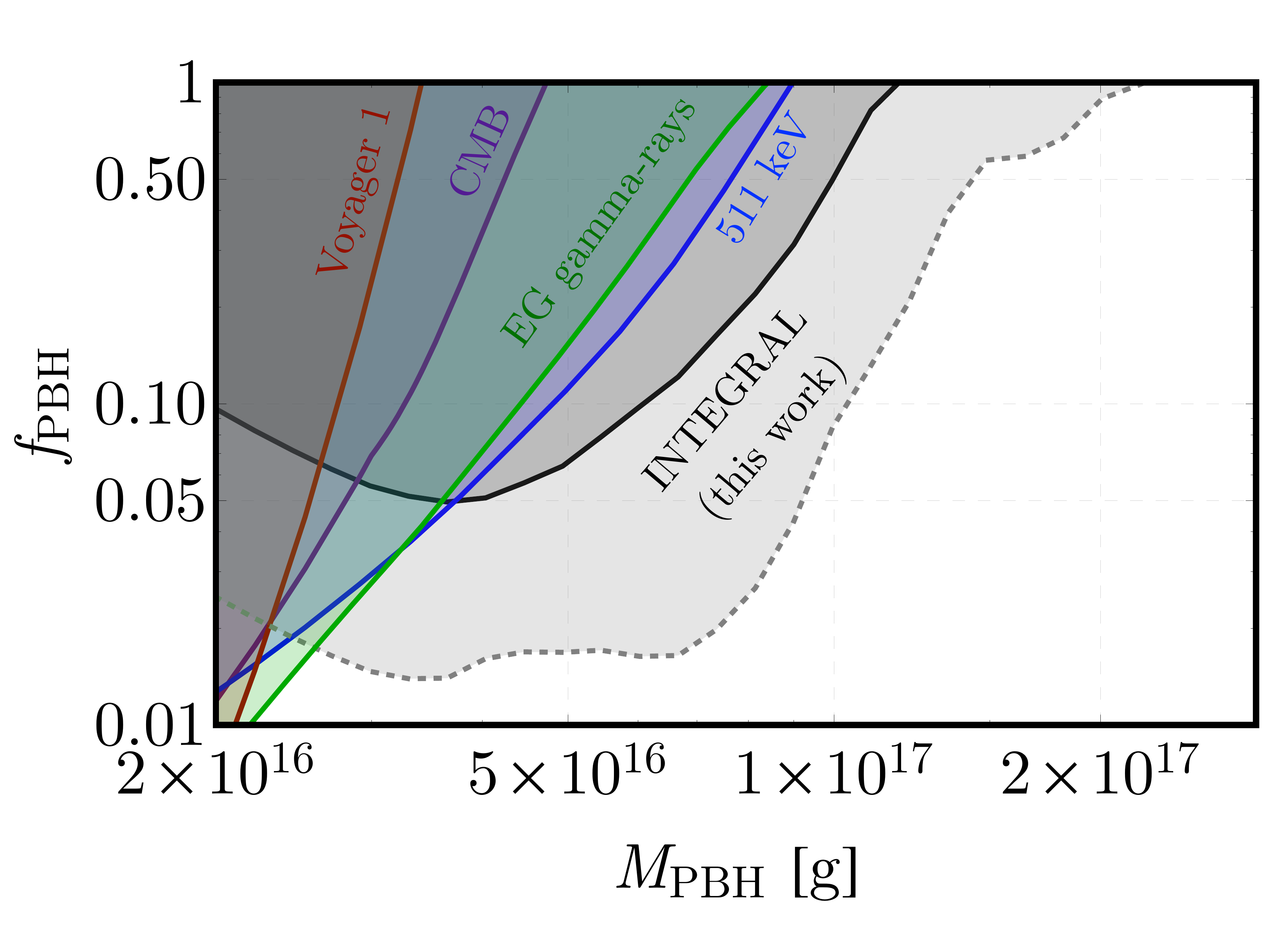}
	\caption{Different constraints on the fraction $f_{\rm PBH}$ of DM that is composed of PBHs. The limit from detection of positrons with Voyager 1 is shown in red~\cite{Boudaud:2018hqb} (propagation model B without background), from the CMB in purple~\cite{Poulter:2019ooo} (varying all parameters), from extra-Galactic gamma-ray emission in green~\cite{Ballesteros:2019exr} (assuming no AGN background), and from the flux of the 511 keV line in the MW in blue~\cite{Laha:2019ssq} (assuming an isothermal DM profile with 1.5 kpc positron annihilation region).
	Our 95\% C.L. constraint, from the Galactic gamma-ray flux measured by INTEGRAL (assuming $\rho_\odot=0.4$ GeV cm$^{-3}$), is shown as the black shaded region.
	We additionally show, in dotted gray, the result that would be obtained with an optimistic analysis of the INTEGRAL data.
	For reference, there are currently no robust constraints to the right of the plot until $M_{\rm PBH}=10^{23}$ g~\cite{Niikura:2017zjd,Katz:2018zrn,Montero-Camacho:2019jte,Smyth:2019whb,Dasgupta:2019cae}.
	}	
	\label{fig:constraintsPBHs}
\end{figure}

We show our limits in Fig.~\ref{fig:constraintsPBHs}, where we see that INTEGRAL can rule out PBHs composing the entirety of the DM for masses up to $M_{\rm PBH}=1.2\times 10^{17}$ g, providing the strongest constraint to date.
Our result improves upon that obtained through the flux of electron-positron pairs in the Galaxy, which would then annihilate and emit a line at 511 keV~\cite{DeRocco:2019fjq,Laha:2019ssq,Dasgupta:2019cae} (this constraint depends on the dark matter profile at the Galactic Center and the propagation distance of low-energy positrons).
Additionally, our constraint is tighter than CMB limits~\cite{Poulter:2019ooo} (see also Ref.\,\cite{Stocker:2018avm} for a similar result), as well as those from Voyager-1 measurements~\cite{Boudaud:2018hqb} (which can vary by more than an order of magnitude depending on the propagation model and background considerations). 
We also obtain stronger constraints using the Galactic gamma-ray emission from PBHs, as opposed to the extra-Galactic component~\cite{Ballesteros:2019exr}.

In Fig.~\ref{fig:constraintsPBHs} we show the most conservative versions of each of these bounds, in order to compare to our robust limits. Modeling of backgrounds and optimistic choices of parameters (for instance in the cosmic-ray propagation) may give rise to nominally stronger but less-robust constraints.
Likewise, our constraints would be strengthened under different assumptions for the analysis; for example, a more optimistic local-DM density of $\rho_\odot=0.6$ GeV cm$^{-3}$~\cite{Buch:2018qdr,Schutz:2017tfp} would yield a tighter constraint of $M_{\rm PBH}>2\times 10^{17}$ g at 68\% C.L. 
In principle, one could also model the astrophysical emission from the Galaxy and subtract it, in order to obtain stronger limits from INTEGRAL data, albeit these would be less robust. 
As a first example, we have found limits using the Galactic-center INTEGRAL spectrum as done for particle DM in Ref.~\cite{Essig:2013goa}. We use the data from Ref.~\cite{Bouchet:2008rp}, which has fine energy resolution (although we remind the reader that this data does not capture every photon, and thus cannot be used to definitively rule out PBHs); as in our main analysis, we simply require that the DM model does not overproduce the 2-$\sigma$ upper limits on the data points. These limits extend further than our robust result; a joint analysis of decaying/annihilating DM, plus other astrophysical sources, using photon data finely binned spatially and in energy, would be optimal for obtaining constraints and could extend the limit to even higher PBH masses

Our results set the strongest lower bound on the PBH mass that is allowed to constitute the entirety of the DM, at $M_{\rm PBH}=1.2\times 10^{17}$ g.
Recent work has cast doubt on PBH constraints due to femtolensing~\cite{Katz:2018zrn}, and capture onto stars~\cite{Montero-Camacho:2019jte}.
Thus, there is a large gap between our result and the next constraint, at $M_{\rm PBH}= 10^{23}$ g, from Subaru microlensing data~\cite{Niikura:2017zjd,Smyth:2019whb}, where PBHs are currently allowed to make up all the DM.  Many ideas have been proposed in order to constrain PBHs in this and higher-mass windows\,\cite{Munoz:2016tmg,Katz:2018zrn,Laha:2018zav,Montero-Camacho:2019jte,Jung:2019fcs, Bai:2018bej,Katz:2019qug,Jung:2017flg,Kuhnel:2019zbc,Zagorac:2019ekv,Kusenko:2020pcg}. 
Additionally, we note that our constraints would be tighter for highly spinning PBHs.
We find, for instance, that for nearly extremal PBHs (with dimensionless spin parameter $a^*=0.9999$) INTEGRAL rules out masses up to $M_{\rm PBH}=10^{18}$ g.

\section{Conclusions}
\label{sec:conclusions}

We have presented constraints on decaying, annihilating, and PBH DM using INTEGRAL measurements of Galactic gamma-ray emission.
We followed a conservative approach, where we use the total measured flux at different Galactic coordinates to set constraints, without assuming any form for the astrophysical contribution.

For decaying DM, we have revisited the constraints from Ref.~\cite{Essig:2013goa}, which we find to be overstated by a factor of $\sim$ $3-10$. Our updated constraints, in Fig.~\ref{fig:tauconst}, are still the strongest for decaying DM masses $m_\chi \in [0.054 - 3.6]$ MeV.
We find similar results to Ref.~\cite{Essig:2013goa} for DM annihilating to electron-positron pairs plus FSR.  
We showed that for DM annihilating to photons, INTEGRAL improves upon the CMB and NuSTAR limits by a factor of $\sim 10$ and $\sim 100$, respectively, for $m_\chi \in [0.027 - 1.8]$ MeV.

We have additionally used the INTEGRAL data to constrain ultralight PBHs.
If these BHs are a component of the cosmological DM, their Hawking radiation will make them appear as gamma-ray point sources, following the DM profile in our Galaxy.
Under conservative assumptions, we showed that PBHs with masses below $M_{\rm PBH}=1.2\times 10^{17}$ g are not allowed to be the entirety of the DM.
While this is an improvement over current constraints, there are still 6 orders of magnitude in mass until microlensing can constrain PBHs~\cite{Smyth:2019whb}.
It is critical to envision new methods to probe this PBH mass window.

\acknowledgements
We wish to thank Andrew Strong and Laurent Bouchet for their help with the INTEGRAL dataset, as well as Samuel McDermott, Kenny C.~Y.~Ng, Anupam Ray, and Kathryn Zurek for discussions and comments on a previous version of this manuscript.
The authors are thankful to the Centro de Ciencias de Benasque Pedro Pascual and the CERN visitor program, without which this collaboration would not have taken place.
RL thanks CERN Theory group for support.
JBM is funded by NSF grant AST-1813694.
TRS is supported by the Office of High Energy Physics of the U.S. Department of Energy under Grant No. DE-SC00012567 and DE-SC0013999.

\bibliography{ULPBH}

\appendix

\begin{figure}[hbtp!]
	\centering
	\includegraphics[width=0.48\textwidth]{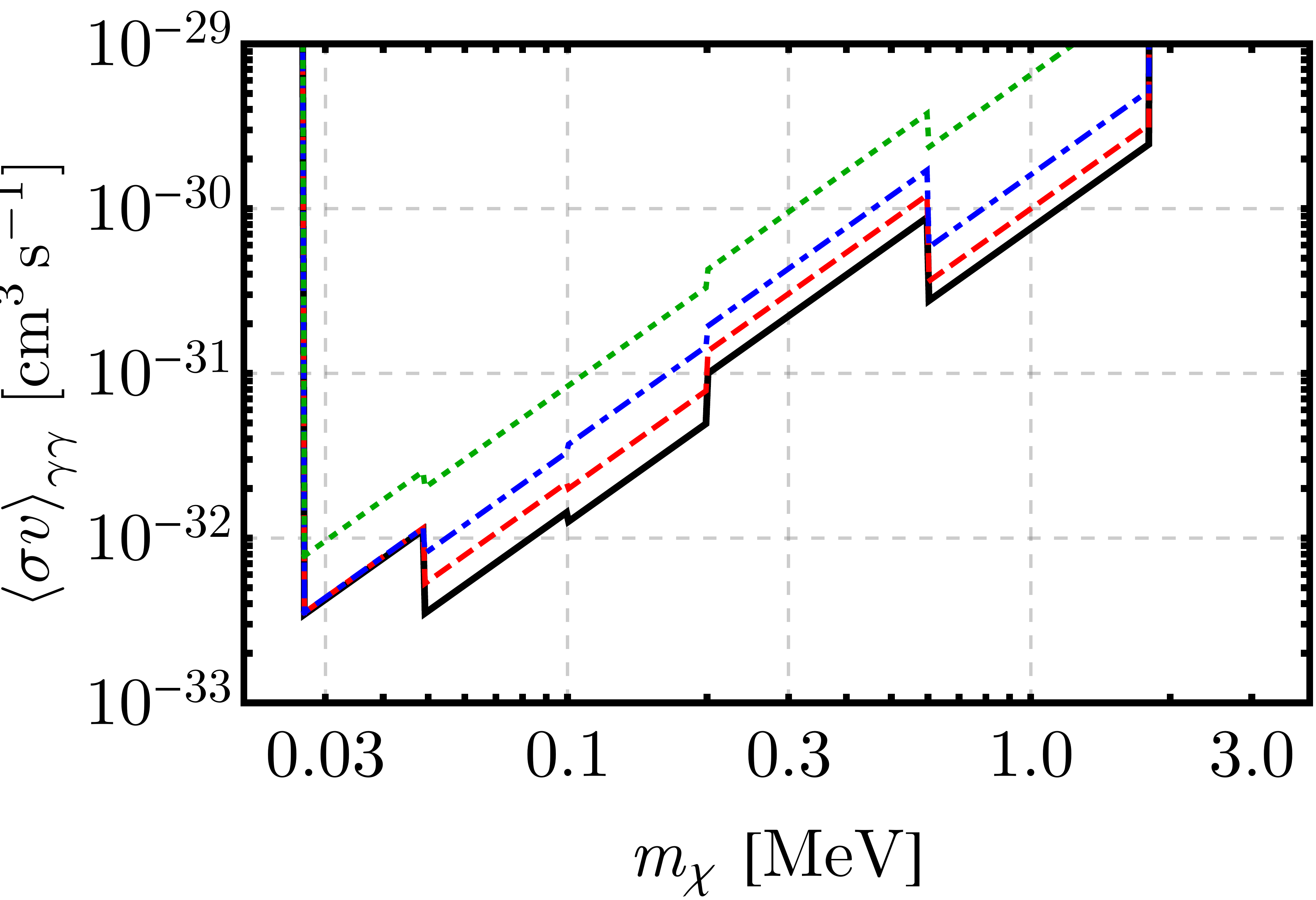}
	\caption{Our INTEGRAL constraints on DM annihilating to two photons, assuming different DM profiles.
	The black line assumes no coring (and is identical for core radii $r_c\leq 0.1$ kpc).
	The three lines assume progressively large cores, of 1 kpc (red dashed), 2 kpc (blue dot-dashed) and 5 kpc (green dotted).
	The constraints do not weaken homogeneously, as which data point sets the constraint changes.
	}
	\label{fig:coreANNDM}
\end{figure}

\section{Impact of the dark-matter profile}
\label{sec:appendix}

Throughout this work we have assumed that the DM in the Galaxy follows a standard Navarro-Frenk-White (NFW) profile.
Here we explore how our results would change if the DM distribution near the Galactic Center was flattened, forming a DM core, which can be a natural consequence of stellar feedback~\cite{Teyssier:2012ie}.
A DM core will alter the $J$ factors, and thus the strength of the constraints we set.
To model that, we will simply flatten the DM distributions in Eq.~\eqref{eq:dJ} to be $\rho(r)=\rho(r_c)$ for $r<r_c$, and the usual NFW result for $r\geq r_c$, where $r$ is the Galacto-centric distance, and $r_c$ is the core radius that we consider.
We will study values of $r_c$ up to 5 kpc, as suggested in Ref.~\cite{Mollitor:2014ara}, although significant cores in MW-sized haloes may be disfavored by more recent simulations~\cite{Chan:2015tna}.
Since the effect of a core is more pronounced for annihilating DM than for decaying DM or PBHs, we divide this Appendix into two parts, studying each case individually.

\subsection{Decaying DM and PBHs}

We first study the change to decaying particle DM and ultralight PBHs.
Note that, both for decaying and annihilating DM, changing the DM profile does not translate into a straightforward rescaling of our limits, since the constraints for each DM mass are set by different energy bands.

We have tested that for both cases there is no appreciable change to our results for $r_c\leq 1$ kpc, as the constraints are driven by data points with relatively large Galactic latitudes ($b\sim 20$ deg.).
Even for larger cores we find only a marginal degradation of the constraints.
For decaying DM we find a 20\% degradation of the $\tau$ limits in the 0.1-1.5 MeV mass range for $r_c=5$ kpc, and no change otherwise (including no change for $r_c\leq 2$ kpc).
For PBHs we likewise find a degradation of 20\% on the $f_{\rm PBH}$ limits, constant over the PBH mass, for $r_c=2-5$ kpc.
Nevertheless, we are able to rule out $f_{\rm PBH}=1$ for $M_{\rm PBH}\leq 1.2\times 10^{17}$ g,
showing that a DM core does not impact our PBH constraints significantly.

\subsection{Annihilating DM}

While the changes for annihilating DM can be more dramatic, we find that for $r_c\leq 0.1$ kpc there is no difference with the standard uncored NFW profile. 
Thus, we will study cases with $r_c=1,$ 2 and 5 kpc.
We show the limits for DM annihilating to photon pairs under the three core assumptions, as well as the standard NFW, in Fig.~\ref{fig:coreANNDM}.
That figure demonstrates that the limits are not just rescaled when changing the DM profile, as they change differently across different masses.

\end{document}